\documentstyle[sprocl,psfig]{article}

\def\be{\begin{equation}}
\def\ee{\end{equation}}
\def\bea{\begin{eqnarray}}
\def\eea{\end{eqnarray}}

\def\h3Mpc{h^{-3}{\rm Mpc}^3 }

\def\h3Mpcinv{h^{3}{\rm Mpc}^{-3} }

\begin{document}
\pagestyle{empty}

\title{GRAVITATIONAL LENSING AND TESTS OF THE COSMOLOGICAL CONSTANT}

\author{B. FORT}
\address{Observatoire de Paris, DEMIRM, 61 avenue de l'Observatoire,\\
75014 Paris, France (fort@mesiom.obspm.fr}

\author{ Y. MELLIER }

\address{Institut d'Astrophysique de Paris, 98 bis boulevard Arago\\
75014
Paris, France (mellier@iap.fr) \\
Observatoire de Paris, DEMIRM, 61 avenue de l'Observatoire,\\
75014 Paris, France}


\maketitle\abstracts{
The case for a  flat Cold Dark Matter model with a positive cosmological
constant $\Lambda$  has been recently  strongly advocated by some 
theoreticians. In
this paper we give the observers  point of view to the light of the
most recent observations with a special emphasis on lensing tests. We
confirm the apparent  cosmic concordance for a flat Universe with  $\Omega_{\Lambda}$ close
to  0.6 but we note that  a low mass density open universe
with no cosmological constant is still quite acceptable for most of the
reliable observational tests, including lensing tests as well.}

\section{Introduction}\label{sec:intro}
A model of the Universe   should be consistent with the  low mass density,
as required by the cosmic evolution of the largest structures and
nucleosynthesis,  the observed value of the Hubble constant $H_0$,  and   an
age   compatible with the observations of globular  clusters and with
 the  ages of
stellar populations in  distant galaxies \cite{Dunlopetal1996}.   With
these  constraints the most likely  region that survives in the space of
cosmological parameters ($H_0$, $\Omega_0$, $\Omega_{\Lambda}$)  seems to force the evidence that we
may live in a flat Cold Dark Matter  Universe with a positive cosmological
constant ($\Lambda$-CDM). It is  important to note that  the  recent measurements
of the angular power spectrum of Cosmic Background Radiation (CBR)
anisotropy are   compatible with $H_0 \simeq 65 \ km/s/Mpc$ and  
$\Omega_{\Lambda} = \Lambda/3 \ H_0^2 \simeq 0.6$
\cite{Lineweaveretal1997}$^{\!,\,}$\cite{Tegmark1996}. Indeed,
 projects
like MAP and Planck Surveyor (2005-2007) that  will allow a  determination
of the values of  the  three cosmological  parameters with an accuracy of
about 10\% \cite{Jungmanetal1996}  or better \cite{Bondetal1997}
should in principle definitely  close one of the most crucial  debate  in
modern astrophysical cosmology.  To follow the striking wording of 
 Ostriker \& Steinhardt 
\cite{OstrikerandSteinhardt1996},  we  have at present a "Cosmic Concordance" of
data and models supporting a flat Universe with 
$\Omega_{\Lambda}  = 0.63 \pm 0.1$ . In fact,
this Cosmic Concordance  was  first  noted about two decades ago 
by Gunn \& Tinsley \cite{GunnandTinsley1975} and since then further re-investigated 
from time to time
to the light of new observational data. More recently,    
 Turner \cite{Turner1997} and
Bagla et al. \cite{Baglaetal1996}  presented   up-dated discussions 
of the problem. They
 emphasize that  there is still now no definitive theoretical arguments against
a non-zero cosmological constant  and that  it is quite possible that  we
have  to  face all the  consequences  of  such a result.

The analysis of the  domain of existence of a  cosmological constant is
possible because for Friedmann-Lemaitre 's models of the Universe  the 3
fundamental cosmological parameters  are linked to the curvature parameter
$\Omega_k = 1-\Omega_0-\Omega_{\Lambda}$,   where  
 $\Omega_0={8 \pi G \rho_0 \over 3 H_0^2}$ is the  density 
parameter expressed with the present-day density of the Universe, $\rho_0$,  
and 
 $\Omega_{\Lambda}$ is the reduced cosmological constant which appears
as an additional term of energy.
   Excluding for the moment the  results of gravitational lensing
tests, let us  summarize briefly  
 the present state  knowledge on  $H_0$ and $\Omega_0$, to the
light of the latest observations.

For the first time,  all  determinations of the value of the Hubble
constant deduced from the HST observations  of Cepheids  are converging to
a value around $63 \pm 10 \ km/s/Mpc$ 
(Sandage \& Tammann 
\cite{SandageandTammann1996} proposed $62 \pm  6 \ km/s/Mpc$). 
Beaulieu et al. \cite{Beaulieuetal1997} have shown with a large sample of
magellanic Cepheids that metallicity effects could  increase  
 this value up to $70 \ km/s/Mpc$. Gouguenheim (this proceeding) presented
   a
comprehensive  discussion of $H_0$, including this problem and the Hipparcos
contribution to the primary stellar distance scale. Moreover, despite an
intrinsic limitation of  lens modeling due to possible  external
gravitational shear or unseen additional masses   on any line of sight, the
time light delay in  the  two multiple quasars  Q 0957+151  and  PG 1115+080
also give the    same result with  about the same error bars 
 \cite{Kundicetal1996}$^{\!,\,}$\cite{Falcoetal1997}$^{\!,\,}$\cite{Keetoneta1996}.    Last but not least,
all others upper bound values  coming from  tests like absolute brightness
of distant supernovae \cite{Wheeleretal1997} and galaxies surface brightness
\cite{Thomsen1997} are fully compatible with this determination.
If we  consider now   that $H_0$ is a least  known with an accuracy of about
15\%, almost all  determinations  of the density parameter  $\Omega_0$
   still  have
large  uncertainties ($0.2 < \Omega_0 < 1$). 
By chance, the number of  independent
observational constraints on $\Omega_0$ is  large, so the
   peak of measurements  around 
 $0.3^{+0.2}_{-0.1}$  obtained from a large
class of tests  seems  significant.  
Let us give a rapid overview of the situation  for 
these major
recent tests. 

Whatever the cosmic scenario of formation of large structures
and galaxies, very high resolution N-body simulations always show that dark
halos can be fitted with  the same radial simple profile only scaled by two
parameters.    Navarro\cite{Navarro1996}  shows that the observed rotation 
curves of
galaxies immersed in such  Universal Dark Halos  can only be explained for
less concentrated halos such as those which  are formed with a  flat
universe    with $\Omega_{\Lambda}= 0.7$. The  study of the redshift 
distribution of damped
Lyman-$\alpha$ halo absorbers has made important progresses during the last
years. It  shows that $\Omega_{\Lambda}= 0$ open-universes with 
$\Omega_0< 0.4$   are inconsistent
with the observations while a flat-universe with $\Omega_{\Lambda}= 0.6$
  or  with $\Omega_0= 1$ \cite{Gardner1997} are acceptable.  Many 
others good  constraints of the
value of $\Omega_0$  come from  the study of masses  content in clusters of
galaxies, the largest (well) observed structures in the Universe. Their
baryon fraction inferred from X-ray studies 
 allows to deduce that $\Omega_0 = 0.32 \pm 0.2$
 \cite{Whiteetal1993} and their
galaxies velocity fields  give imply that $\Omega_0 = 0.3 \pm 0.1$
 \cite{Carlbergetal1997},  whereas the direct lensing measurements of 
their total
gravitational masses within a radius of 0.5 Mpc gives  
 $\Omega_0$ in the range =
0.2-0.5 (cf Mellier et al, this conference). With the weak lensing
method, higher lensing densities  corresponding to  $\Omega_0$ close to 1  
cannot
be  excluded   from some tentative  measurements of  extremely weak
gravitational shear  in volume of several Mpc-sizes around cluster
centers (Mellier et al., this proceeding).  But 
for such observations  the correction of instrumental
distortion may not yet be fully under 
control \cite{bm95}$^{\!,\,}$\cite{kaiseretal95}. Surprisingly the  lensing
results obtained  for large scale structures  seems  comparable to those
coming from  the latest interpretation of large cosmic  
flows  \cite{Deckel1997}. Note that the lensing method  seems more 
reliable and is  able to
progress significantly in the near future. Therefore, the most likely value
for $\Omega_0$ emerging  from  all these  observational constraints  seems equal to
$0.3^{+0.2}_{-0.1}$.

At least, one crucial tests on  $\Omega_{\Lambda}$
 for  flat models is now emerging  from
the observations of high-z supernovae. It allows in
principle the determination of the acceleration 
parameter $q_o= \Omega/2-\Omega_{\Lambda}/2$ with an 
accuracy of $\pm$0.2 as soon 
as we will be able to detect
and observe a sample of SNs at $z > 1$ \cite{GoodbarandPerlmutter1995}. So far,
with  a sample of SNs only at  $z < 0.4$ 
Perlmutter et al. \cite{Pelmutteretal1996}   note that
 $q_0$  could be  actually in the 0-0.5 interval.  
Similarly, Bender et al. \cite{Benderetal} recently  demonstrate  that we 
can actually  use cluster ellipticals
as cosmological candles.  With two  clusters at $z=0.375$ they
tentatively found $q_0$  in the range 0-0.7. It is clear  that such  new
observational tests  do not much favor the Cosmic Concordance paradigm.
They  can progress a lot if suitable telescope time is given to these
programs  and we are  expecting a lot from them for a direct measurement of
$q_0$.

\section{Lensing test}
Hence, to the light of all these  results,  it becomes   interesting to
look at an other set of  almost independent tests on $\Omega_{\Lambda}$: the 
lensing tests. The most classical  ones concern the frequency of 
multiply imaged quasars,
that in fact probes the volume of the Universe per unit redshift at large
distance, or the probability distribution of the redshifts and the
distribution of the separation of the multiple  lensed images  which depend
also  on $\Omega_{\Lambda}$. Many works  were published on these tests 
and we  just refer
in the following to a few recent papers.  Some recent tests  try to use
the cosmic evolution of condensations of mass on large scales that can be
studied with gravitational  arcs or weak lensing modeling, or  to probe
directly the curvature parameter  of the  Universe  $\Omega_k$.   These  
tests
use the  measurements of  deviations of light rays passing through
same lenses but coming from distant sources at different redshifts.
 Deflexion angles induced by lenses
 are  proportional to  the  effective angular 
distance $D = D_{ol}.D_{ls}/D_{os}$ (see Mellier et al. this conference). 
The $D_{ij}$ are respectively
the angular distance between the observer,  the lenses and  the 
sources, and $D_{ij}$  is
defined by
\be
H_0 \ D_{ij} = {c \over (1+z_j) } \vert \alpha_0 \vert^{-1/2}
\int_{1+z_1}^{1+z_2} \ \left(2 \sigma_0 y^3+\alpha_0 y^2 + \lambda
\right)^{-1/2} dy,
\ee
where $sinn(x)$ is respectively $sin (x)$ for close universe, $sinh (x)$
 for an
open universe  and  $x$ for a flat universe, and
\be
\sigma_0={8 \pi G \over 3 H_0} \ , \ \ \lambda=\sigma_0-q_0 \ , \ \ 
\alpha_0=1+q_0-3\sigma_0 \ , \ \ k=-{H_0^2 R_0^2 \over c^2} \alpha_0 \ .
\ee
Hence,  deviation angles, impact
parameter like Einstein radius or location of critical lines depend on  the
existence of $\Omega_{\Lambda}$.

In practice direct geometrical lensing tests with $D$, in particular for
flat universes as predicted by inflation,  work  only when the cosmological
constant  becomes larger than about 0.6 because it is only above this value
that the $\Lambda$ effect  becomes observable at large $z$.  The observational
difficulty comes from the small excess of deviation, which is  never
larger than 2-5\% of the deviation expected for the 
$\Omega_{\Lambda}=0$ case having the same
$\Omega_0$. This is why this new class of tests are only relevant 
 for  lensing-clusters with large
angular scales. In the following we  summarize the  results and new trends
for all the class of lensing  tests.

\subsection{Statistics of lensed quasars}
The volume per unit redshift at large distance  and consequently  the
relative number of lensed  sources  for a fixed co-moving density
increases with  increasing  cosmological constant. This can be used to set
constrain on $\Omega_0$  for an open-universe (say $\Omega_0> 0.15$). 
 After the pioneer work of Turner et a. \cite{tog}, 
Turner \cite{Turner1990} 
 used this effect  and constrained $\Omega_{\Lambda}$
   for flat-universe models from the statistics of  lensed quasars. Many successive
authors applied the method and Kochanek \cite{Kochanek1996}  
used  it  with the most
recent  optical or radio (fair) sample of lensed quasars  including a detailed
study of all the sources of errors: lens modeling, quasar counts, galaxy
lenses counts and dust effects at large redshift.  At the same time, he
also considered his earlier  original approach  that uses the probability
distribution of the lenses redshift. For a non-zero cosmological constant
this probability distribution reaches a maximum at a redshift which increases
with $\Omega_{\Lambda}$. The sensitivity of the second  method is 
 smaller but it gives
consistent results for each  selected  quasar surveys . The Kochanek's paper
is  certainly now a reference  for this kind of  work. It  basically
confirms  previous results but it gives a lower and reliable  upper limit
$\Omega_{\Lambda}< 0.65$
  at the 2-$\sigma$ confidence level for a flat cosmology. This limit
is now very  difficult to disprove.  It seems that only  a strong dust
obscuration, hundred times larger  than  currently admitted, can relax a
bit this limit.

\subsection{Statistics on quasar pair separation}
When the  model (velocity dispersion, core radius, etc..) and lens redshift
of multiple high-$z$ 
 quasars lenses  are known the  mean splitting  of  the images  is
sensitive to $\Omega_{\Lambda}$ \cite{Paczynskietal1981}.  A recent 
application of the
statistics of quasar  separation  was made by 
Myungshin et al. \cite{Myungshinetal1996} who
selected seven "isolated" elliptical lenses with  reliable observational
data to constrain $\Omega_0$ and $\Omega_{\Lambda}$.  Using a maximum 
likelihood method on the
probability to observe simultaneously the photometric and geometric
characteristics of the seven lensing events, they excluded  high mass density
universe and  in particular a flat-universe   with $\Omega_0=1$   at the 98\%
confidence level. They also found that  a flat-universe with  
$\Omega_{\Lambda} = 0.65 \pm 0.25$ (about 50\% confidence level)  
is more  acceptable than  low-mass open
universes with  $\Omega_0 <  0.3$  and $\Omega_{\Lambda}=0$.  This method 
depends  critically  on
the total number of such lensed quasars events and it is probable that this
number will not increase   much during the next decade.  However this
method could inspire the work which is done with gravitational arcs.

\subsection{Gravitational arc(let)s}
Up to now, several tens of multiply imaged distant galaxies  have  already
been detected in clusters of galaxies and this number will probably
increase by a large factor within the next years. These gravitational arcs
provide a first sample that could be tentatively used for  tests on the
existence of the cosmological constant.  Assuming that the  mean redshift
of the blue arcs is about $z =1$, Wu and Mao \cite{WuandMao1996}  
noted that their number
is  indeed enhanced  if the cosmological constant is large. But they
confirm  the earlier results of Bartelmann \cite{bart95} that  irregular  distribution of masses increase also the
probability to find an arc so that no claims can be done 
on $\Omega_{\Lambda}=0$ without a
detailed modeling  of clusters from strong and weak lensing. Besides,
Wilson et al. \cite{Wilsonetal1996}  propose to constrain  
cosmological constants from the
cosmic evolution of condensations of masses probed by lens modeling.
Simulations shows that the method  could be efficient to discriminate an
Einstein-de Sitter $\Omega_0$ universe from any  low mass density universe but
they are almost insensitive to  discriminate a flat universe with a
cosmological constant $\Omega_{\Lambda}< 0.7$
 from the corresponding  $\Omega_{\Lambda}=0$ universe.
The same result is  obtained analytically  for  interpretations of the
"cosmic shear" both in the linear and  non linear regime (see Mellier et
al, this conference). This  comes from the fact that even for very distant
sources the lensing probability  by  any foreground condensation of mass
is maximum at redshift  in the range 0.3-0.6. Hence  the lensing effect
mainly probe an intermediate-redshift Universe  which mostly depends 
on  $\Omega_0$
and the power spectrum  of initial density fluctuation $P(k)$.

 Good modeling of  several   multiple  arcs within a same cluster  having
different  redshifts  range like 0.5-1 and 3-5 could give constraints on
the the value of the cosmological constant \cite{FortandMellier1994}. As an
example, for an isothermal lens potential,  as soon as the velocity
dispersion is known, the radial position of the critical lines where arcs
at a given redshift are formed depends only upon the effective angular
distance $D(z,\Omega_0,\Omega_{\Lambda})$ defined above. The  position of 
the most distant arcs
depends on $\Omega_{\Lambda}$  after the modeling has fitted the low 
redshift arcs. The
effect of the cosmological constant is then  detectable for 
$\Omega_{\Lambda} > 0.5$  on  a
few arcseconds scale.   Actual  cluster potentials  are by far more 
 complex that these models which miss small scale 
structures,  but the outstanding 
 images obtained with HST images permit to  considerably improve 
 detailed investigations  because 
   the critical line where multiple
arcs and  unresolved pairs merge can be located with high accuracy. 
 Thus, good models that control  the various potential gradients 
 on an arcminute scale can be done from strong and
weak lensing.
 Maximum likelihood methods should in principle
be applied as soon as such  fair lensing events will be observed.  Since
the work of Mellier et al. \cite{mellieretal93}, the
possibility of doing  almost perfect modeling of arcs systems  is  
demonstrated and has been used in numerous lensing-clusters observed with 
HST (see the most recent splendid examples  given by  
Seitz et al. \cite{Seitzetal1997}, and also  by Kneib et al
\cite{kneibetal96} in A2218).
 In summary,  for a given lens model, the method  compares the
location of a low-redshift critical line to that of a high-redshift one. So
far,  no cases where  found where  the modeling of two arc system at very
different redshifts is completely reliable. Most of the clusters with a
simple potential turns out to have  only low redshift arcs systems because the
search for very  distant arcs or unresolved pairs ($z>3$)   has to be done
mainly with HST or IR observations which are still lacking!

In view of this limitation, Fort et al. \cite{Fortetal1997} have recently 
proposed to
find the location of distant critical lines ($z>3$) by using the radial
magnification effect  of clusters \cite{Broadhurst1996}.   With very deep
images in cluster Cl0024+1654 and A370 they showed that the number density
of galaxies in the very faint magnitude bin $B = 26-28$ and $I = 24-25.5$
 displays an (anti)magnification bias  with a depletion that extends to a
radial distance of about 1 arcminute, as compared to 32 arcseconds for the
giant arc radius. They  suggest that the selected  bins of $B$-galaxies have
a redshift distribution
of the objects  with two peaks respectively at $z =1$ (60\%) and $z=3$ (40\%) ,
and that the selected bin of  $I$  galaxies  allows to approximate the
location of  a high-redshift critical line ($z>3.5-4$) very close from the
last critical line at  infinite redshift. Using the modeling of the lenses
already published they  found that the location of this high redshift
critical line rather favor a flat cosmology with $\Omega_{\Lambda}$
   larger than 0.6.
This  method can be significantly improved with a large  sample of  arcs 
clusters particularly by using
a maximum  likelihood methods   applied to the probabilities of reproducing
their observed local shears and convergences  (equivalent to the
magnification bias). A strong improvement can come from the new 
possibility to
accurately select various bins of source redshifts using  their color
indices \cite{brunneretal97}.

\section{Conclusions}
If we  accept  that $H_0$  is around $60 \ km/s/Mpc$
 and that the Universe is old
as it is inferred from the theory of the  evolution of stars and stellar
populations in very distant galaxies, it is   impossible to  have a flat
 matter-dominated Universe $\Omega_0 =1$. In fact, many other 
observational tests
including lensing tests also  exclude this possibility except for a few
that are not  yet  fully reliable because of possible observational bias
or controversial theoretical hypothesis. It is true that  many tests on
 the cosmological constant   yield  apparently concordant  
results  that are in favor of a  flat Universe 
with $\Omega_0 = 0.3^{+0.2}_{-0.1}$ and $\Omega_{\Lambda} = 0.6 \pm 0.2$. 
This is  compatible  with the  reliable  upper limit 
$\Omega_{\Lambda} < 0.65$ derived by
Kochanek from quasars statistics.  However,  for most of the
tests, open universes  having   $\Omega_0 \simeq 0.3$  and 
$\Omega_{\Lambda}  = 0$ are also often
acceptable.  If we are actually living in a universe with a cosmological
constant close to the Kochanek limit  it is worth to improve the lensing
tests particularly with the gravitational arcs.
  It is  sure that the question will be  also quickly addressed
with the on-going large distant supernova surveys because the $q_0$ test has a
remarkable efficiency. There is no doubt that the search for a
non-zero cosmological constant corresponds to a high priority observing
program  without waiting for the MAP and Planck Surveyor surveys. At
present and from the observational point of view it is still too early to
make a case for a non-zero cosmological constant. More reliable
observational facts have to be accumulated.

\section*{Acknowledgments}
We thank F. Bernardeau, P. Schneider, S. Seitz and L. van Waerbeke
 for  fruitful discussions
and  enthusiastic  collaborations.  We thank J. Lequeux for  
his comments on the manuscript.

\end{document}